# Capacitance Performance of Sub-2-nm Graphene Nanochannels in Aqueous Electrolyte

Yinghua Qiu, and Yunfei Chen*

*School of Mechanical Engineering and Jiangsu Key Laboratory for Design and Manufacture of Micro-Nano Biomedical Instruments, Southeast University, Nanjing, 211189, Jiangsu province, People's Republic of China*

*Corresponding author: *yunfeichen@seu.edu.cn*
 Tel:138-158-888-16
 Fax:086-25-52090504

*Abstract:*

Molecular dynamics simulations were used to explain the origin and properties of electrical double-layer capacitance in short graphene nanochannels with width below 2 nm. The results explain the previously reported experimental result on the non-monotonic dependence of the capacitance with the channel width. The mechanism for the anomalous increase of the capacitance in sub-1-nm in pore diameter is attributed here to the width-dependent radial location of counterions in the nanochannels, and the restricted number of co-ions. Decrease of the channel width lowers the number of co-ions and positions the counterions closer to the channel walls. For nanochannels with width ranging from 1 to 2 nm, co-ions are allowed to enter the nanochannel, and both types of ions assume alternating layered distributions leading to the decrease of the capacitance. Voltage is another control parameter which allows understanding capacitance in graphene nanochannels. As the voltage increases, due to limited space near the charged surface, more counterions need to be located in the center of the nanochannel resulting in further capacitance decrease.

# Key words:

Electric double layer capacitor, graphene nanochannels, molecular dynamics simulation, and

energy storage

# Introduction

As the availability of fossil fuels decreases, alternative energy sources, such as solar, wind and other renewable clean power, need to be exploited to satisfy society's demand. Unfortunately, places abundant with such resources are usually far away from the human habitat. Thus, efficient energy storage devices are becoming crucial to resolve the shortage of energy. In the past decade, electric double layer capacitors (EDLCs) have been widely researched as a promising electric energy storage system, also named supercapacitors or ultracapacitors.[1-2] They can store the charge at the electrode/electrolyte interface through revisable electrostatic adsorption of ions under a voltage applied on the electrodes.[3] Cations from the electrolyte accumulate at the negatively charged surfaces, and anions at the positively charged ones. Both electrodes are usually highly porous to provide the largest possible access area to the ions.[4] The electrolyte used to supply charged ions can be aqueous electrolytes,[5] organic electrolytes,[6] or ionic liquid[7-8] each possessing a different upper limit for applied potentials and ionic mobility.[9] Due to the high ionic mobility and purely physical adsorption to the charged surfaces, EDLCs have higher power density and more extraordinary durability than batteries,[10] which allow them have broad application as portable energy storage devices, such as the power sources of hybrid electric vehicles and camera flash bulbs, which need huge transient currents.[11]

However, the energy density of EDLCs is typically less than 10 Wh/kg,[12] making it the primary limitation of the wide-scale usage of EDLCs. Because the capacitance on electrodes is proportional to the effective surface area that ions can approach[3], the energy stored in EDLCs has a linear relationship with the surface area. In order to raise the energy density of the EDLCs, a variety of carbon-based materials with high specific surface area have been considered for electrode designs[13], such as carbon nanotubes,[14-15] graphene-based composites[16-18], and other kinds of carbon based materials.[19-20] Usually, pores smaller than the size of solvated electrolyte ions are recognized as incapable of contributing to charge storage.[21] However, using carbide-derived carbon, Chmiola et al. discovered the double layer capacitance in an organic electrolyte showed anomalous increase with average pore size less than 1.0 nm, which is smaller than the size of the solvated ion. In a different experiment, the area normalized capacitance of activated carbon electrodes immersed in 6.0 M KOH solution has been found by Lota et al.[5] to increase from 6.4 μF/cm$^2$ to 11.6 μF/cm$^2$ with the mean pore diameter decreasing from 14.5 to 10.6 Å.

Because the experiment can only provide the actual capacitance performance of electrodes immersed in electrolytes, the nanoscale characteristics at the electrode-electrolyte interfaces have widely been investigated by molecular dynamics (MD) simulation and other simulation methods[22]. The discovered mechanism of the capacitance variation with nanopore size is of great importance to facilitate the development of better-performance EDLCs. Similar anomalous increase of capacitance with decreasing nanopore size was also obtained by simulations. The specific capacitance normalized to the pore surface area is reported by Shim et al.[23] to increase as the carbon nanotube diameter decreases from 2.0 nm to 0.9 nm in room temperature ionic liquid. Wu et al.[24] found that the capacitance of slit-shaped nanopores in contact with ionic liquids increases with the pore width decreasing from 0.91 nm to 0.75 nm. Interestingly, supercapacitor capacitance exhibits oscillatory behavior with a decaying envelope as the nanopore size increases using classical density functional theory. Wu et al. thought it should be attributed to the interference of two electrical double layers near each slit wall. Additionally, the effects of electrode surface roughness,[25] solvent,[6] curvature of nanopore[26] and the temperature[27] on the capacitance of EDLCs have been well researched. The dynamics of the ions,[28] ion distribution[23, 29] and liquid structures[30] have also been studied to help explain the performance of EDLCs.

In the theoretical aspect, the capacitance of EDL can be treated as the combination of the capacitance from two regions: the Stern layer capacitance and the diffusion region capacitance.[3] For a slit-shaped EDLC, its capacitance can be described by the new sandwich capacitance model raised by Feng et al.[12] that considers the size of ions in the parallel-plate model. For the cylindrical nanopore EDLCs, the prediction of the capacitance is complex. Huang et al.[31] took the curvature of the pore into consideration, and divided the capacitive behavior into two different types based on the size of nanopore. The electric double-cylinder capacitor model and electric wire-in-cylinder capacitor model are used to calculate the capacitance in the pores larger than 2 nm and smaller than 1 nm, respectively.

Due to its higher ionic mobility, aqueous electrolytes have huge potential to improve the power density of the supercapacitors. Although this anomalous behavior of capacitance scaling has been ascribed to the distortion of the solvation shell of ions inside nanopores in organic electrolyte[21], the underlying physics of the capacitance of microporous materials remains poorly understood, especially for aqueous electrolyte and when the pore size is between 1 and 2 nm. Some recent simulations investigated the ion and water distributions in slit-shaped nanopores using aqueous electrolyte. Feng et al.[12] studied the distribution of $K^+$ ions in electrified slit-shaped nanopores with size ranging from 9.36 to 14.7 Å using MD simulations. They found that the counterion distribution differs qualitatively from that described by Poisson-Boltzmann theory. However, in this work they didn't take co-ions into consideration, which showed an important influence on the capacitance when the pore size smaller than 1 nm. Using MD simulation, the charge storage in 0.78 nm slit-shaped pores was investigated by Wu et al.[32] Their results showed that with the increase of surface charge, the capacitance increased due to the removal of co-ions from the pore and entrance of counterions. When all co-ions are expelled from the pore, the capacitance reached a maximum. Kalluri et al.[33] studied the partition and structure of aqueous NaCl electrolyte ~1.8 M in ionic strength in carbon-slit electrodes from 0.65 nm to 1.6 nm. They found that the maximum partition coefficient obtained at −0.15 $C/m^2$ is 0.65nm for $Na^+$ ions. However, the system they used was too small to consider two capacitors with different surface charges. The co-ion effect was almost fully inhibited due to the attraction reduced by the other oppositely charged electrodes.

So, we conducted a MD simulation to investigate the ionic structure and capacitance in negatively charged graphene nanochannels immersed in aqueous NaCl solution. As previously reported,[5] large ionic concentrations are needed to achieve higher energy densities of EDLCs, thus in our model the concentration we selected was 4.0 M to provide large number of $Na^+$ ions. A bulk-nanochannel-bulk model was used and the system could reach electrical neutrality at its potential minimum through the free diffusion of the counter ions and co-ions. In the simulation model, the applied surface potential was conducted with a certain surface charge density. The natural ionic distributions which are not influenced by artificial neutrality can be obtained.

# Molecular dynamics model

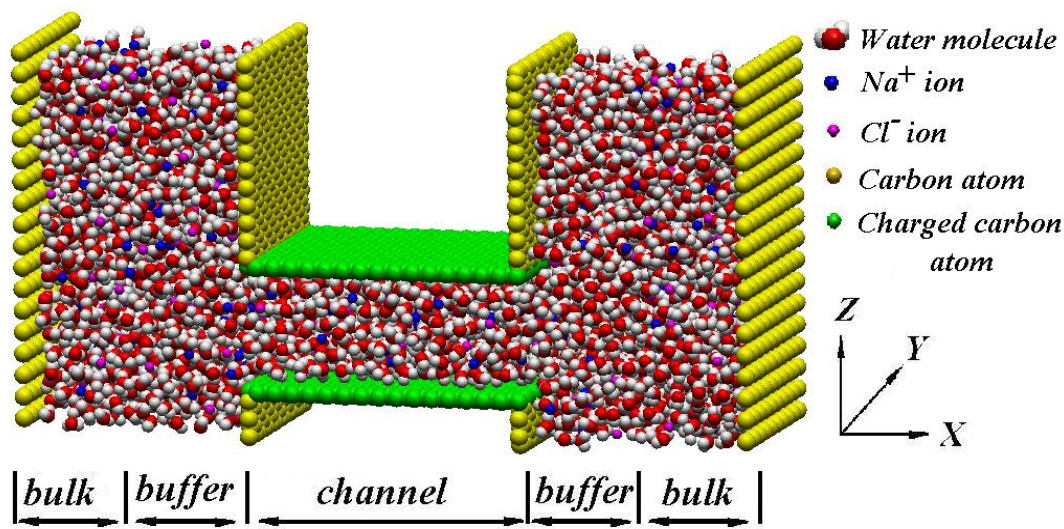

Fig. 1 Schematic diagram of the MD model. The dimensions of the system in x, y and z directions are 10.1, 2.84 and 5.0 nm, respectively. The channel length is 3.81 nm.

The schematic diagram of the bulk-nanochannel-bulk model used in the simulations is exhibited in Fig. 1, which is assumed to be infinite in the *y* direction using periodic boundary conditions. There are a bulk and buffer regions on each side of the model. Outside the bulk region, there is a layer of sparse carbon atoms constructed as a solid boundary to the aqueous solution. The bulk region is filled with a predefined salt solution to provide enough ions and water molecules. The buffer region is used to judge whether the system achieves equilibrium. In our simulation, once the salt concentration in the buffer region is not changed with the simulation time and remains at a predefined concentration for a sufficient period of time, the system is believed to reach equilibrium. The following simulations are used to get the final results. The middle part of the model is the nanochannel that we want to simulate and where get the final results from. Both the upper and the bottom solid walls are composed of a single graphene layer. During the simulation the wall atoms are frozen without thermal vibration. The surface charge density is set by uniformly distributing a certain number of charges to the carbon atoms in the graphene layers. Eight nanochannels with different width were selected to be investigated. The widths were 1.75, 1.5, 1.25, 1.125, 1.0, 0.875, 0.75 and 0.5 nm.

In the simulations, the surface charge density was set as −0.15 C/m$^2$. At the beginning of each simulation, pure water was placed in the regions adjacent to the channel openings, marked as buffer in Fig. 1, and in the channel; the bulk regions were filled with 4.0 M NaCl.[34] Once the simulation started, ions would diffuse from the bulk regions to the nanochannel until the concentration gradients disappeared. The concentration gradients were evaluated from the buffer regions. Before the system reached equilibrium, the solution in the two bulk regions was replaced periodically to assure its concentration. With two bulk regions, more ions could be supplied, allowing the system to reach equilibrium much faster. In all cases, the water molecules and ions numbers are listed in Table 1.

Table 1. The numbers of the water molecules and ions in the MD systems.

| channel width (nm) | number of water molecules | number of $Na^+$ ions | number of $Cl^-$ ions |
| --- | --- | --- | --- |
| 1.75 | 3468 | 244 | 224 |
| 1.5 | 3337 | 237 | 217 |
| 1.25 | 3241 | 231 | 211 |
| 1.125 | 3218 | 219 | 199 |
| 1.0 | 3131 | 229 | 209 |
| 0.875 | 3112 | 215 | 195 |
| 0.75 | 3022 | 213 | 193 |
| 0.5 | 2887 | 210 | 190 |

Our simulations were conducted using the codes developed by our lab.[34-36] In the simulations, the TIP4P model[37] was selected to simulate the water molecules and the SETTLE algorithm[38] was chosen to maintain the water geometry. The Lennard-Jones (LJ) potential was used to describe the interactions between different atoms, except hydrogen-X pairs (X is the atom species in the system) and carbon-carbon pairs. The parameters of simulations were taken from ref.39. The electrostatic interactions among ions, water molecules and surfaces charges were modeled by the Ewald summation algorithm.[40] The motion equations were integrated by the leap-frog algorithm with time step of 2.0 fs. The solution system was maintained at 298 K by Berendesen thermostat.[41] The cut-off used for the short range interaction is 1.2 nm, and the relaxation time of the thermostat is 100 fs. The first run lasting 8 ns was used to equilibrate the system. Another 6-ns-long run followed to gather the statistical quantities.

# Results and discussions

The counterion and co-ion distributions perpendicular to the surfaces that obtained from graphene nanochannels with different widths are plotted in Fig. 2. We can find that the counterions and co-ions show an alternating layer distribution when the channel width larger than 1nm, which is similar to the results in organic solutions.[7] Comparing the three cases of 1.25, 1.5 and 1.75 nm, the peak locations of ions seems stable. $Na^+$ ions mainly accumulate in a layer located at 0.46 nm away from the surfaces and its first peak appears at 0.27 nm, which is close to the hydration radius[42] of the $Na^+$ ions and affected by the hydrophobic property of the surface.[43] As the width of the channel decreases, the numbers of $Na^+$ and $Cl^-$ ion aggregation peaks shrink. When the pore size is 1.25 nm, there are only two accumulation $Na^+$ ion layers, located near both the channel surfaces due to the increased constriction in the nanochannel. Because of the electrostatic repulsion provided by the surface charges, $Cl^-$ ions can only appear in the region farther than 0.63 nm away from the surface where the surface potential has been nearly screened by the $Na^+$ ions. Therein this region, $Cl^-$ ion concentration increases rapidly and exceeds that of $Na^+$ ion, which causes the so-called charge inversion[44-45] that cannot be explained by the mean field theory Poisson-Boltzmann equation.[46] In the case of 1.75 nm high channel, the concentration of $Cl^-$ ions exceeds that of $Na^+$ ions three times. As the channel width decreases, it becomes harder for $Cl^-$ ions to enter the channels, becoming nearly impossible when the channel width drops to 1.125 nm. When the pore size dips below 1 nm, $Cl^-$ ions can hardly enter the pore. The channel can only accommodate $Na^+$ ions, which mainly locate in the first accumulation layer near the surface as the pore gets smaller. This is caused by the disappearance of the oscillatory water distribution[47] between both hydrophobic graphene walls that keeps $Na^+$ ions far away from the surfaces. The variation of counterion distribution in our work is different from other MD simulation results.[12, 48] In our MD model, the simulated systems reach electric neutrality[33] through the free diffusion of the cations and anions without any artificial effects. So, the obtained ion distributions can reflect the practical physical model with channel size in nanoscale.

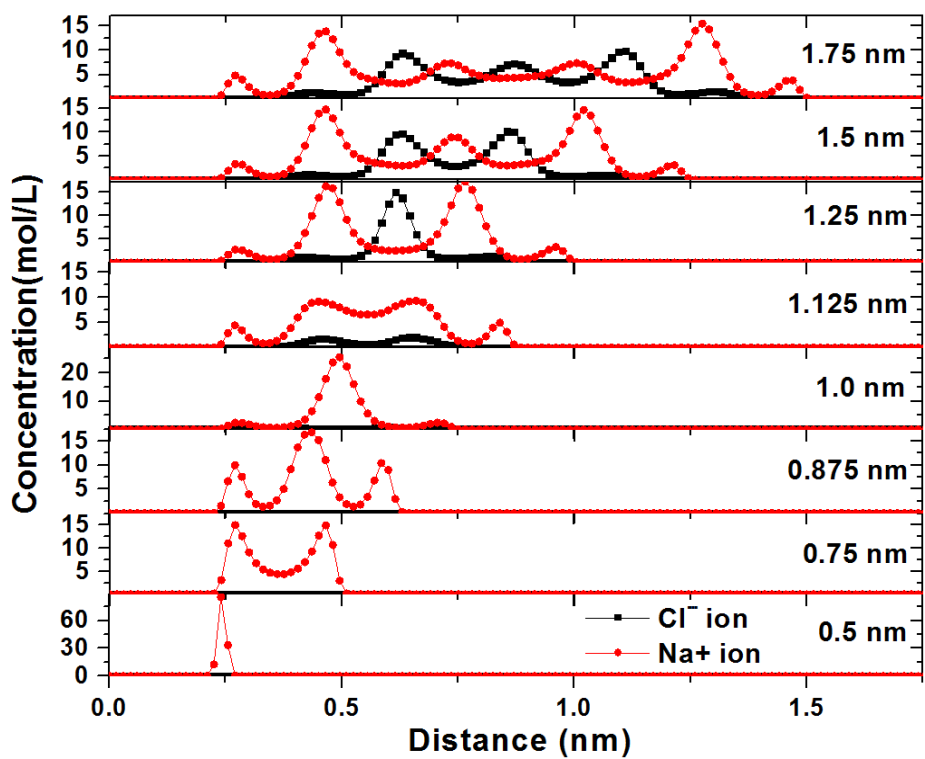

Fig. 2 The ion distributions in the nanochannels with width from 1.75 nm to 0.5 nm. The surface charge densities in the eight MD models are the same as −0.15C/m$^2$.

The alternating layer distribution of cations and anions in organic electrolyte near uniformly charged surface is mainly due to the appearance of charge inversion affected by the large volume of the ions.[49] However, in aqueous solution, the ions are usually so small that charge inversion rarely occurs near a uniformly charged surface. In our case, due to the restricted space in nanochannels, the confinement is much stronger. In addition to the interface effect, the aggregated water molecules in specific layers have huge influence on the locations of the counterion peaks, which mainly appear in the valleys of water distributions (Fig.S1).[48, 50]

During the simulation process, Cl$^-$ ions mainly accumulated in the center of the nanopores, where they were not strongly influenced by the surface charges and had more hydrated water molecules. With the decrease of the pore size, the space that Cl$^-$ ions can reach is compressed, which leads to less and less Cl$^-$ ions entering the pore. In Fig.3, the numbers of ions and the percentage $P$ of the Na$^+$ ion in the nanochannels with different width are shown: $P = N_{Na^+} / N$ where, $N_{Na^+}$ is the number of Na$^+$ ions and $N$ is the total number of ions in the channel regions.

In the 1.75 nm case, there is enough space to accommodate many ions. Thus, the percentage of Na$^+$ ions is about 60%, a little larger than 50%. With the decrease of the pore size, it is much harder for Cl$^-$ ions to enter the nanochannels. The percentage of Na$^+$ ions shows almost a nonlinear increase and reaches its maximum 100% in the 0.875 nm width channel. When the width was dropped to 0.4 nm, both cations and anions hardly entered the pore.

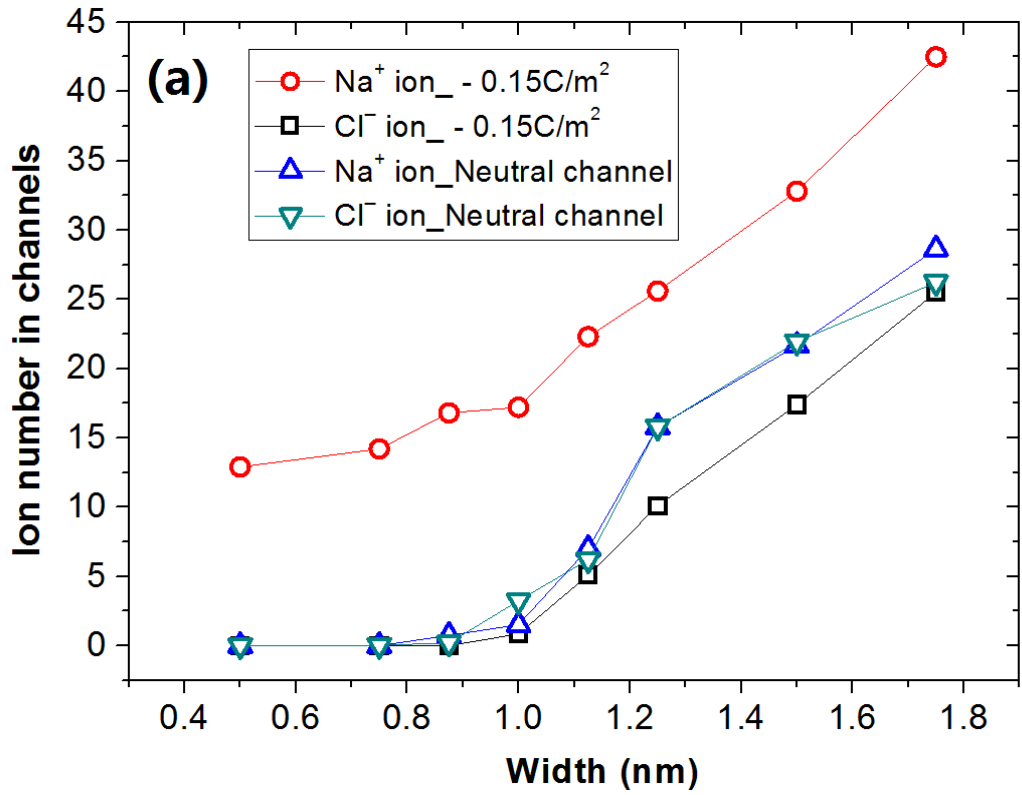

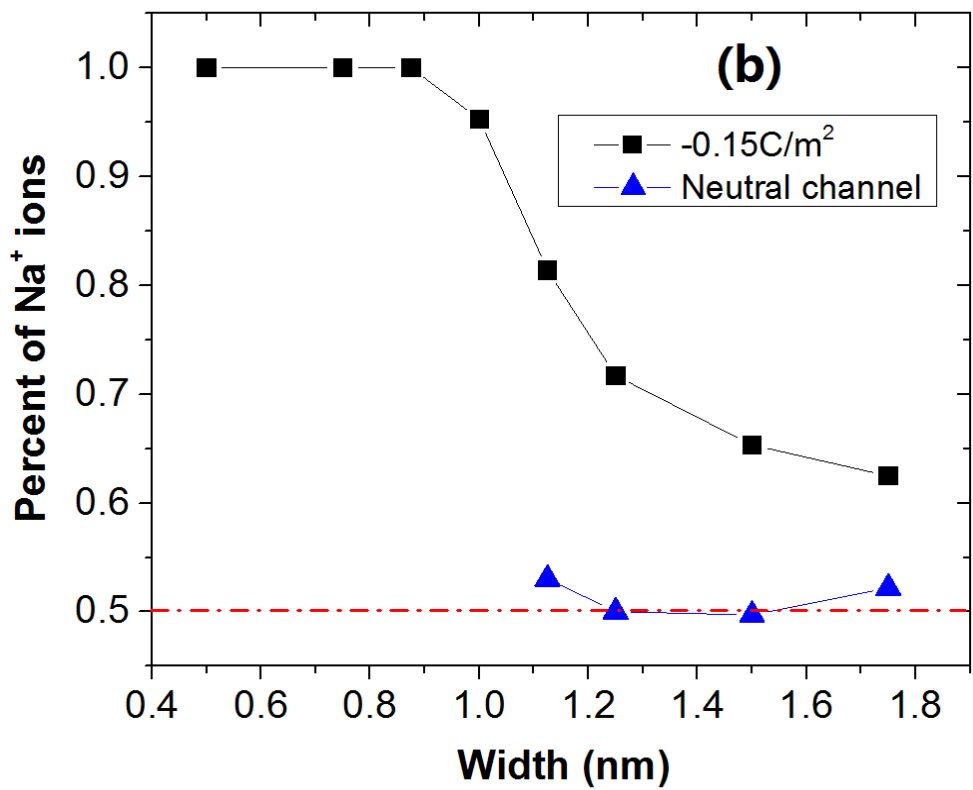

Fig. 3 (a)The numbers of ions in the charged and neutral channels with different widths. (b)The percent composition of Na$^+$ ions in the charged and neutral nanochannels with different widths.

The coordination numbers of Na$^+$ ions[28] were analyzed to explore their hydration performance in the nanochannels. In Fig. 4 (a) the average coordination numbers in the channels with different sizes are exhibited. It is found the average coordination numbers in the first hydrated shell of ions (0.35 nm) in the nanochannels with width from 1.75 nm to 0.875 nm are nearly the same, about 6 water molecules.[28] The width has little influence on the hydration performance of ions confined in nanochannels. This is mainly because the largest accumulating layer counter ions that have the same hydration performance in different high channels take the predominate part in amount. However, when the width reduces further, the hydration number of Na$^+$ ions decreases sharply, which is caused by the destruction of the oscillatory water layers confined between the channel surfaces (Fig.S2). The coordination number distributions with distance from the surface are described in Fig. 4(b). The counterions in the center of the channels usually have more water molecules nearby, which can also provide a hindrance for their approaching to the surface. The coordination number in the bulk solution is 6.88.

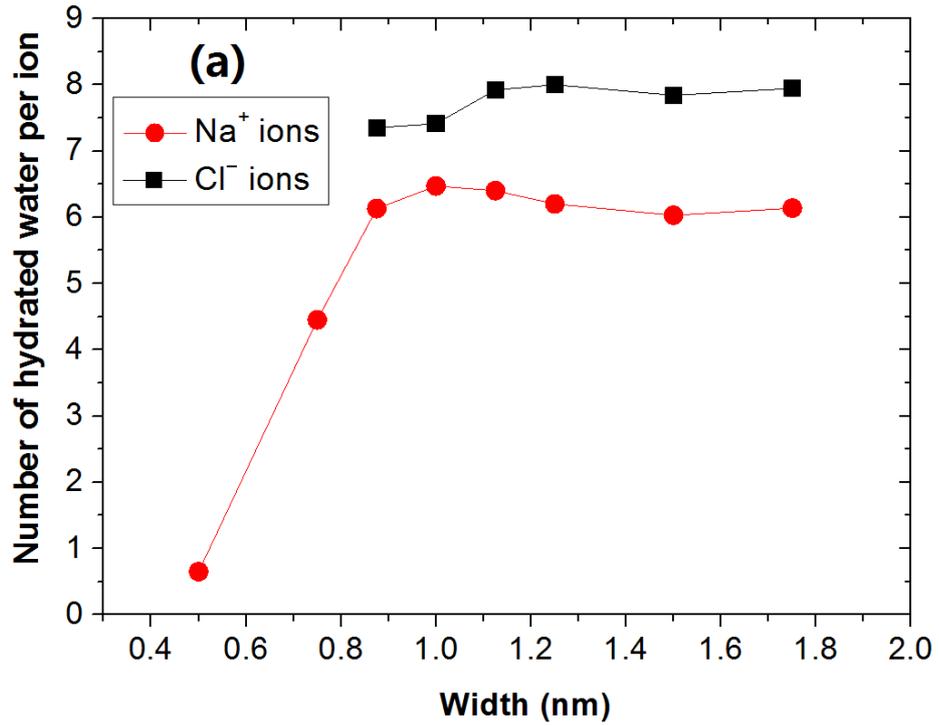

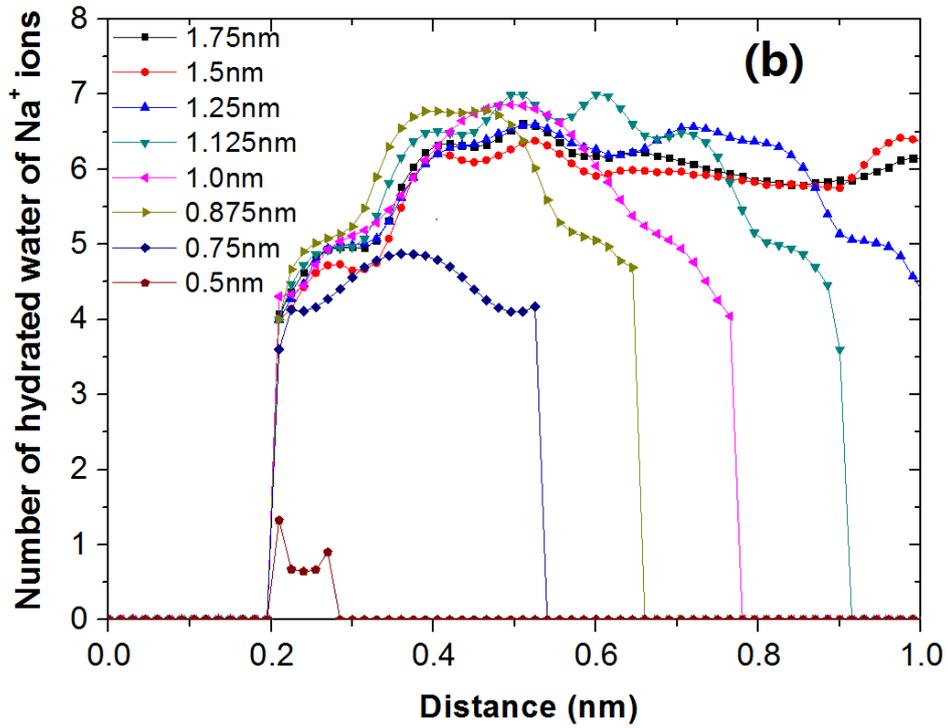

Fig. 4 (a) The number of hydrated water per ion in the nanochannels with different widths. (b) The hydrated water number distributions of Na$^+$ ions perpendicular to the surface. Note that the asymmetry for the smallest channel might stem from limited number of water and ions in the channel.

Based on the ion distributions in the nanochannels with different widths, the surface potentials were calculated.[51] Then using the equation $C = \sigma/(\varphi - \varphi_0)$,[52-53] the capacitance was obtained (Table S1 and S2), where σ is the surface charge density, φ is the charged surface potential, and $\varphi_0$ is the surface potential when the surface is uncharged. The surface potentials of uncharged graphene walls were obtained from the ion distributions in neutral systems. The obtained capacitances with channel width are exhibited in Fig. 5(a), which is very close to the experimental data[5] with corresponding pore sizes. The theory prediction is based on the double layer capacitance when the width is larger than 1 nm using $C = \dfrac{\varepsilon_r \varepsilon_0}{d-a}$, while that of sub-1 nm nanochannels are obtained from the equation $C = \dfrac{\varepsilon_r \varepsilon_0}{(H/2)-a}$ in ref.12. In the equations, $C$ is the capacitance, $\varepsilon_0$ the permittivity of a vacuum, $d$ is the location of the main accumulating layer of ions, and $H$ is the width of the nanochannel, as well as a=0.095 nm[46], $\varepsilon_r$=3.33[12] are used as the radius of the bare ions and the electrolyte dielectric constant, respectively. It is found that our MD simulation data agreed well with the theoretical prediction in the sub-1 nm nanochannels. As the pore size decreases, the capacitance of the EDL increases due to the nearer proximity of the counterion layers. With the continual decrease of the pore size, the hydrated ions may have larger size than the pore. Once the nanopore is reduced too small to accommodate the counterions, the capacitance will decrease with the continued decrease of the pore size due to the difficulty for $Na^+$ ion access. So, it is predicted that, the capacitance of the EDL can reach its maximum for a nanopore with a certain size. In the channels with width from 1.125 nm and 1.75 nm, the predicted capacitance is constant due to the stable structure of $Na^+$ accumulated layers. However, the MD simulation results show a different trend from the theoretical prediction, which indicates a reduction in capacitance as channel width decreases from 1.75 to 1.125 nm. We think the character of EDLCs in aqueous electrolyte is mainly due to the net ionic charge in the nanochannel, which also has close relationship with the ion structure. Usually, the co-ions in nanochannels[32] are neglected in the earlier works[12,33]. From the alternative layer distribution of the counterions and co-ions shown in Fig.2, the charge inversion becomes more obvious as the channel width shrinks from 1.75 nm to 1.25 nm. So, the corresponding capacitance shows a decrease trend.

In a negatively charged nanochannel, the energy is stored by the accumulation of counterions.

Fig. 5(b) shows the ratio of the net ionic charges to the channel width. With the channel width shrinking, this ratio increases, which means the porous electrodes with smaller pore size can accommodate more counterions, not only due to the enhancing surface area but also due to the smaller occupying space of counterions. In our case, if the graphene electrodes have uniform nanochannels with 0.5 nm in width, they should have the largest energy density.

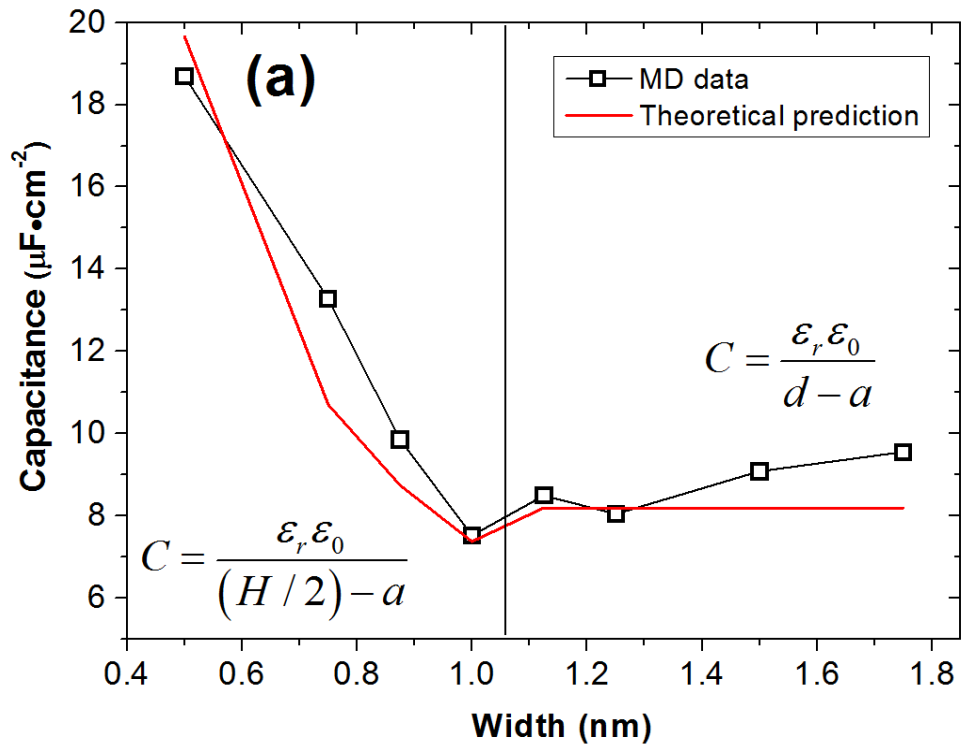

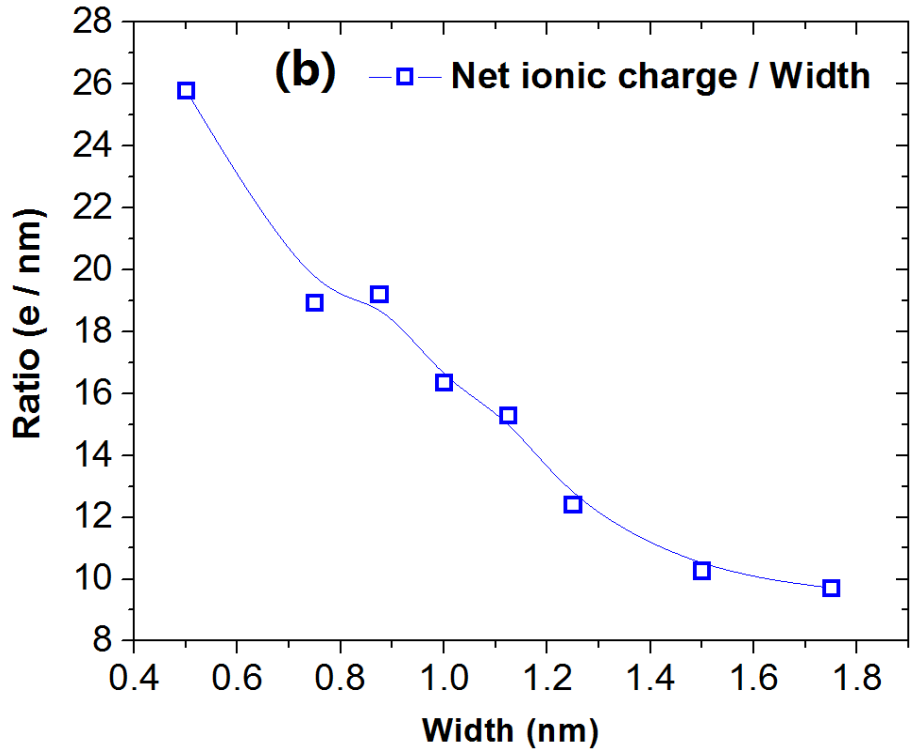

Fig. 5(a) The comparison between the capacitance obtained from MD simulations with theoretical prediction. (Table S1) (b) The ratio of the net ionic charges to the channel width. The net ionic charge is obtained by subtracting the number of Cl⁻ ion from that of Na+ ion in the channels with different widths.

In Fig.2, as the channel width decrease from 1.0 nm to 0.875 nm, the counterions appear to approach the surfaces. In order to figure out the reason for this phenomenon, another four cases with the same width but different surface charge densities from −0.05 to −0.125 $C/m^2$ were studied. The ion distributions and capacitance are shown in Fig.6 and Fig. 7, respectively.

From Fig. 6(a), $Cl^-$ ions can hardly enter the channels and $Na^+$ ions show layering structure. The number of $Na^+$ ions increases linearly with surface charge density shown in Fig. 6(b). When the graphene walls were charged at −0.05 $C/m^2$, two $Na^+$ ion layers with small peak appear at 0.3 nm away from each wall surface. As the surface charge density increases, much more counterions are attracted in the nanochannels to screen the surface charges. The peaks of the first accumulated layers near each wall surface become larger and saturate as the surface charge density high enough. Due to the limited space in the channels, many counterions cannot approach the surface, but remain in the center of the channels. It can be found from Fig.6(a), when the surface charge density reaches −0.1 C/m², many Na⁺ ions accumulated in the channel center and formed another counterion layer. This layer eventually becomes the main counterion accumulation peak when the surface charge density increases to −0.15 C/m².

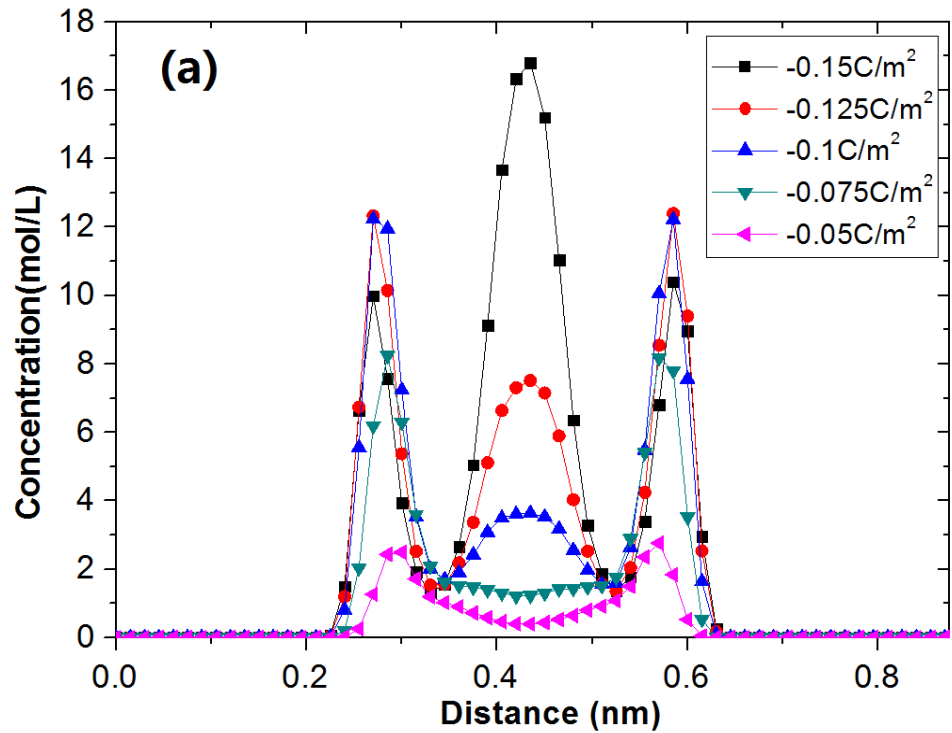

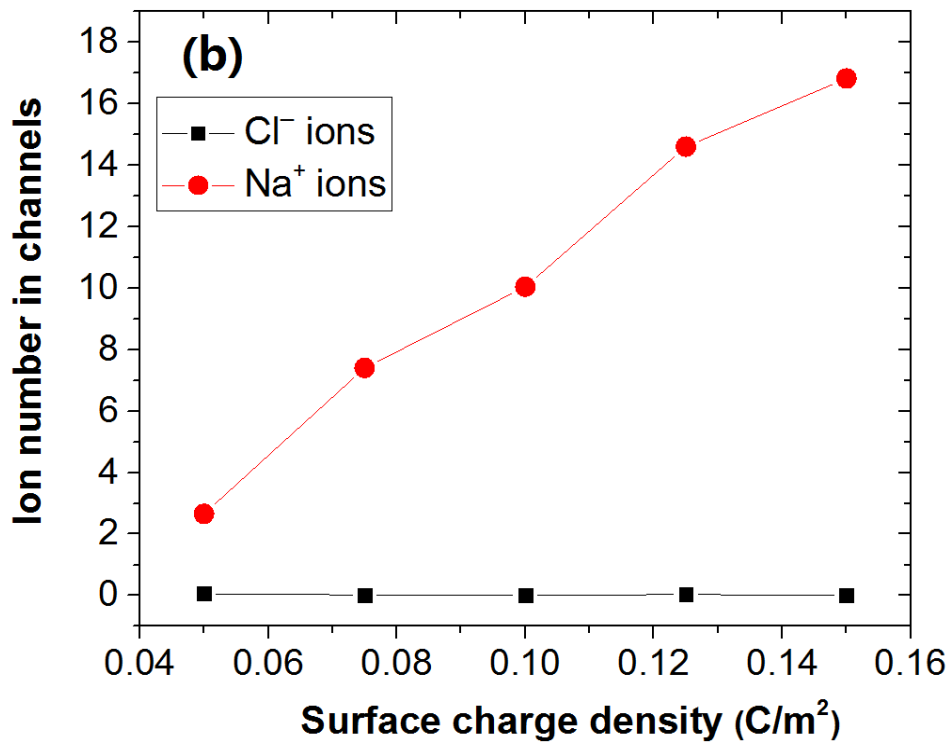

Fig. 6(a) Ion distributions in the nanochannels with surface charge density from −0.05 to − 0.15C/m$^2$ with 0.875 nm in width. (b)The numbers of ions in the nanochannels with different surface charge densities.

The capacitance performance of the graphene nanochannels with the same width but different surface charge densities is shown in Fig.7. With the surface charge density increasing, the capacitance decreases which shows the same trend as other works.[25]. From the ion distributions in Fig. 6(a), we think this may be caused by the farther main location of the counterions. When the surface charge density becomes higher, the main counterions preferred to aggregate in the center of the nanochannels i.e. farther away from the surfaces due to the limited space near the surface. Because the effective distance between the charged surface and the main location of counterions increases and the surface charges cannot be screened so well compared to that in weakly charged channels, the capacitance decreases as the surface charge density increases.

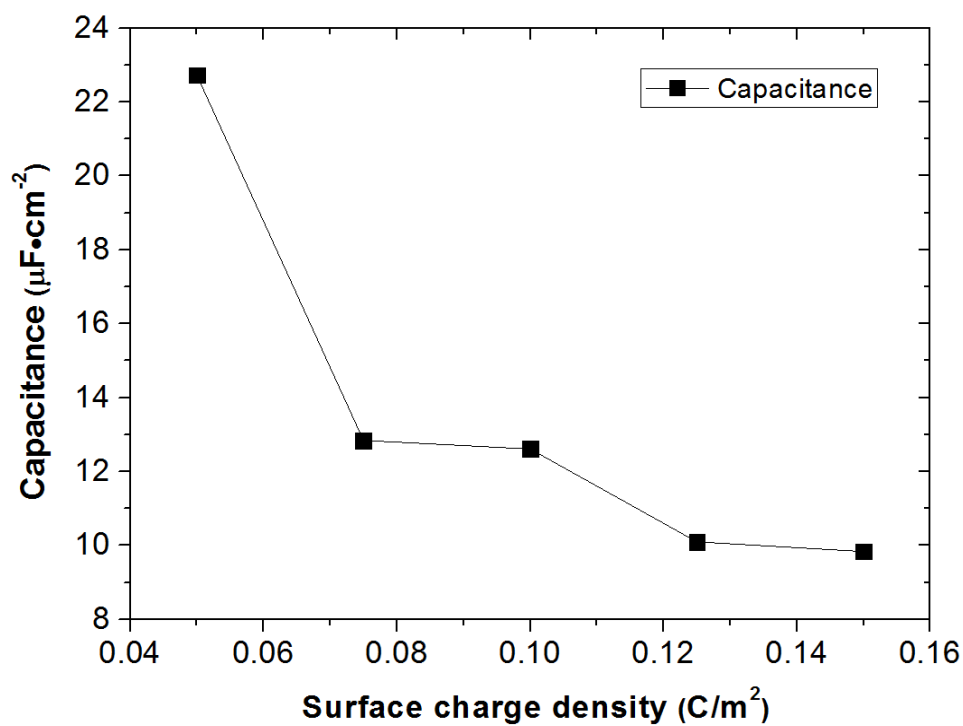

Fig. 7 The capacitance obtained from 0.875 nm nanochannels with different surface charge densities. (Table S2)

# Conclusion

Electrical double layer capacitors have attracted a great deal of interest and the efforts have been focused on increasing their energy density. Because the EDLC energy density is directly correlated with the area normalized capacitance, it is of great importance to study the performance of nanoporous capacitors with sizes smaller than 2 nm i.e. the approximate diameter of the solvated ions. Through MD simulation, we showed that the EDLCs capacitance shows a non-trivial dependence on the nanochannel width with the largest capacitance for 0.5 nm. From the obtained distributions of ions and water in the nanochannels with size larger than 1 nm, we found that the performance of the EDLCs is mainly determined by the amount of co-ions, which control the degree of charge inversion at the solid-liquid interfaces. As the channel size decreases below 1 nm, no co-ions are able to enter the channels and the capacitance increases due to proximity of counterions to the charged surfaces. We concluded and proposed here that a nanopore with the size that can only accommodate the counterions will reach the maximum capacitance and highest energy density. Our results therefore provide guidelines to the experimental groups to prepare nanochannel system which is perfectly ion selective. It is indeed possible that other factors, e.g. interactions between dipole moment of a solvent and ions in a solution might lead to more complex dependence of capacitance as recently shown in ref. 54. Our future studies will include more parameters, such as solvent dipole moment and type of ions for the capacitance performance of the nanochannel.

# Acknowledgement

The authors acknowledge financial support from the Natural Science Foundation of China (Grant numbers 51435003 and 51445007). This work was partly supported by the Fundamental Research Funds for the Central Universities, the Innovative Project for Graduate Students of Jiangsu Province (Grant No. CXZZ13_0087), the Scientific Research Foundation of Graduate School of Southeast University (YBJJ 1322) and the China Scholarship Council (CSC 201406090034). We thank Prof. Z. Siwy and T. Plett in the University of California at Irvine for carefully reading and improving this manuscript.

Supporting Information Available: The water molecules and ions distributions in 1.75 nm high nanochannel with surface charge density −0.15 C/m$^2$ (Figure S1). The water molecules and ions distributions in 1.125, 0.875, 0.75 and 0.5 nm high nanochannel with surface charge density −0.15 C/m$^2$ (Figure S2). The surface potentials of the charged and neutral channels. (Table S1). The surface potentials of the 0.875 nm in width channels with different surface charge densities. (Table S2) This material is available free of charge via the Internet at http://pubs.acs.org.

# References

1. Conway, B., *Electrochemical Supercapacitors: Scientific Fundamentals and Technological Applications*; Kluwer Academic/Plenum: New York, 1999.

2. Frackowiak, E., Carbon Materials for Supercapacitor Application. *Phys. Chem. Chem. Phys.* **2007**, *9*, 1774-1785.

3. Burt, R.; Birkett, G.; Zhao, X. S., A Review of Molecular Modelling of Electric Double Layer Capacitors. *Phys. Chem. Chem. Phys.* **2014**, *16*, 6519-6538.

4. Pohlmann, S.; Lobato, B.; Centeno, T. A.; Balducci, A., The Influence of Pore Size and Surface Area of Activated Carbons on the Performance of Ionic Liquid Based Supercapacitors. *Phys. Chem. Chem. Phys.* **2013**, *15*, 17287-17294.

5. Lota, G.; Centeno, T. A.; Frackowiak, E.; Stoeckli, F., Improvement of the Structural and Chemical Properties of a Commercial Activated Carbon for Its Application in Electrochemical Capacitors. *Electrochim. Acta* **2008**, *53*, 2210-2216.

6. Jiang, D.; Jin, Z.; Henderson, D.; Wu, J., Solvent Effect on the Pore-Size Dependence of an Organic Electrolyte Supercapacitor. *J. Phys. Chem. Lett.* **2012**, *3*, 1727-1731.

7. Jiang, D.; Jin, Z.; Wu, J., Oscillation of Capacitance inside Nanopores. *Nano Lett.* **2011**, *11*, 5373-5377.

8. Fedorov, M. V.; Kornyshev, A. A., Ionic Liquids at Electrified Interfaces. *Chem. Rev.* **2014**, *114*, 2978-3036.

9. Lewandowski, A.; Olejniczak, A.; Galinski, M.; Stepniak, I., Performance of Carbon-Carbon Supercapacitors Based on Organic, Aqueous and Ionic Liquid Electrolytes. *J. Power Sources* **2010**, *195*, 5814-5819.

10. Simon, P.; Gogotsi, Y., Materials for Electrochemical Capacitors. *Nat. Mater.* **2008**, *7*, 845-854.

11. Miller, J. R.; Simon, P., Materials Science - Electrochemical Capacitors for Energy Management. *Science* **2008**, *321*, 651-652.

12. Feng, G.; Qiao, R.; Huang, J.; Sumpter, B. G.; Meunier, V., Ion Distribution in Electrified Micropores and Its Role in the Anomalous Enhancement of Capacitance. *ACS Nano* **2010**, *4*, 2382-2390.

13. Wang, G.; Zhang, L.; Zhang, J., A Review of Electrode Materials for Electrochemical Supercapacitors. *Chem. Soc. Rev.* **2012**, *41*, 797-828.

14. Li, X.; Rong, J.; Wei, B., Electrochemical Behavior of Single-Walled Carbon Nanotube Supercapacitors under Compressive Stress. *ACS Nano* **2010**, *4*, 6039-6049.

15. Pandey, S.; Maiti, U. N.; Palanisamy, K.; Nikolaev, P.; Arepalli, S., Ultrasonicated Double Wall Carbon Nanotubes for Enhanced Electric Double Layer Capacitance. *Appl. Phys. Lett.* **2014**, *104*, 233902.

16. Raccichini, R.; Varzi, A.; Passerini, S.; Scrosati, B., The Role of Graphene for Electrochemical Energy Storage. *Nat. Mater.* **2015**, *14*, 271-279.

17. Zhu, Y.; Murali, S.; Stoller, M. D.; Ganesh, K. J.; Cai, W.; Ferreira, P. J.; Pirkle, A.; Wallace, R. M.; Cychosz, K. A.; Thommes, M., et al., Carbon-Based Supercapacitors Produced by Activation of Graphene. *Science* **2011**, *332*, 1537-1541.

18. Yoo, J. J.; Balakrishnan, K.; Huang, J.; Meunier, V.; Sumpter, B. G.; Srivastava, A.; Conway, M.; Mohana Reddy, A. L.; Yu, J.; Vajtai, R., et al., Ultrathin Planar Graphene Supercapacitors. *Nano Lett.* **2011**, *11*, 1423-1427.

19. Pech, D.; Brunet, M.; Durou, H.; Huang, P.; Mochalin, V.; Gogotsi, Y.; Taberna, P.-L.; Simon, P., Ultrahigh-Power Micrometre-Sized Supercapacitors Based on Onion-Like Carbon. *Nat. Nanotechnol.* **2010**, *5*, 651-654.

20. Gogotsi, Y.; Nikitin, A.; Ye, H. H.; Zhou, W.; Fischer, J. E.; Yi, B.; Foley, H. C.; Barsoum, M. W., Nanoporous Carbide-Derived Carbon with Tunable Pore Size. *Nat. Mater.* **2003**, *2*, 591-594.

21. Chmiola, J.; Yushin, G.; Gogotsi, Y.; Portet, C.; Simon, P.; Taberna, P. L., Anomalous Increase in Carbon Capacitance at Pore Sizes Less Than 1 Nanometer. *Science* **2006**, *313*, 1760-1763.

22. Merlet, C.; Rotenberg, B.; Madden, P. A.; Taberna, P.-L.; Simon, P.; Gogotsi, Y.; Salanne, M., On the Molecular Origin of Supercapacitance in Nanoporous Carbon Electrodes. *Nat. Mater.* **2012**, *11*, 306-310.

23. Shim, Y.; Kim, H. J., Nanoporous Carbon Supercapacitors in an Ionic Liquid: A Computer Simulation Study. *ACS Nano* **2010**, *4*, 2345-2355.

24. Wu, P.; Huang, J.; Meunier, V.; Sumpter, B. G.; Qiao, R., Complex Capacitance Scaling in Ionic Liquids-Filled Nanopores. *ACS Nano* **2011**, *5*, 9044-9051.

25. Vatamanu, J.; Cao, L.; Borodin, O.; Bedrov, D.; Smith, G. D., On the Influence of Surface Topography on the Electric Double Layer Structure and Differential Capacitance of Graphite/Ionic Liquid Interfaces. *J. Phys. Chem. Lett.* **2011**, *2*, 2267-2272.

26. Feng, G.; Jiang, D.; Cummings, P. T., Curvature Effect on the Capacitance of Electric Double Layers at Ionic Liquid/Onion-Like Carbon Interfaces. *J. Chem. Theory Comput.* **2012**, *8*, 1058-1063.

27. Lockett, V.; Sedev, R.; Ralston, J.; Horne, M.; Rodopoulos, T., Differential Capacitance of the Electrical Double Layer in Imidazolium-Based Ionic Liquids: Influence of Potential, Cation Size, and Temperature. *J. Phys. Chem. C* **2008**, *112*, 7486-7495.

28. Feng, G.; Qiao, R.; Huang, J.; Sumpter, B. G.; Meunier, V., Atomistic Insight on the Charging Energetics in Subnanometer Pore Supercapacitors. *J. Phys. Chem. C* **2010**, *114*, 18012-18016.

29. Wander, M. C. F.; Shuford, K. L., Alkali Halide Interfacial Behavior in a Sequence of Charged Slit Pores. *J. Phys. Chem. C* **2011**, *115*, 23610-23619.

30. Smith, A. M.; Lovelock, K. R. J.; Gosvami, N. N.; Licence, P.; Dolan, A.; Welton, T.; Perkin, S., Monolayer to Bilayer Structural Transition in Confined Pyrrolidinium-Based Ionic Liquids. *J. Phys. Chem. Lett.* **2013**, 378-382.

31. Huang, J.; Sumpter, B. G.; Meunier, V., Theoretical Model for Nanoporous Carbon Supercapacitors. *Angew. Chem. Int. Ed.* **2008**, *47*, 520-524.

32. Wu, P.; Huang, J.; Meunier, V.; Sumpter, B. G.; Qiao, R., Voltage Dependent Charge Storage Modes and Capacity in Subnanometer Pores. *J. Phys. Chem. Lett.* **2012**, *3*, 1732-1737.

33. Kalluri, R. K.; Ho, T. A.; Biener, J.; Biener, M. M.; Striolo, A., Partition and Structure of Aqueous Nacl and Cacl2 Electrolytes in Carbon-Slit Electrodes. *J. Phys. Chem. C* **2013**, *117*, 13609-13619.

34. Qiu, Y.; Chen, Y., Counterions and Water Molecules in Charged Silicon Nanochannels: The Influence of Surface Charge Discreteness. *Mol. Simulat.* **2014**, *41*, 1187-1192.

35. Chen, Y.; Ni, Z.; Wang, G.; Xu, D.; Li, Electroosmotic Flow in Nanotubes with High Surface Charge Densities. *Nano Lett.* **2008**, *8*, 42-48.

36. Ge, Y.; Xu, D.; Yang, J.; Chen, Y.; Li, D., Ionic Current through a Nanopore Three Nanometers in Diameter. *Phys. Rev. E* **2009**, *80*, 021918.

37. Jorgensen, W. L.; Chandrasekhar, J.; Madura, J. D.; Impey, R. W.; Klein, M. L., Comparison of Simple Potential Functions for Simulating Liquid Water. *J. Chem. Phys.* **1983**, *79*, 926-935.

38. Miyamoto, S.; Kollman, P. A., Settle - an Analytical Version of the Shake and Rattle Algorithm for Rigid Water Models. *J. Comput. Chem.* **1992**, *13*, 952-962.

39. Cohen-Tanugi, D.; Grossman, J. C., Water Desalination across Nanoporous Graphene. *Nano Lett.* **2012**, *12*, 3602-3608.

40. Yeh, I. C.; Berkowitz, M. L., Ewald Summation for Systems with Slab Geometry. *J. Chem. Phys.* **1999**, *111*, 3155-3162.

41. Berendsen, H. J. C.; Grigera, J. R.; Straatsma, T. P., The Missing Term in Effective Pair Potentials. *J. Phys. Chem.* **1987**, *91*, 6269-6271.

42. Tansel, B.; Sager, J.; Rector, T.; Garland, J.; Strayer, R. F.; Levine, L. F.; Roberts, M.; Hummerick, M.; Bauer, J., Significance of Hydrated Radius and Hydration Shells on Ionic Permeability During Nanofiltration in Dead End and Cross Flow Modes. *Sep. Purif. Technol.* **2006**, *51*, 40-47.

43. Joseph, S.; Aluru, N. R., Why Are Carbon Nanotubes Fast Transporters of Water? *Nano Lett.* **2008**, *8*, 452-458.

44. Grosberg, A. Y.; Nguyen, T. T.; Shklovskii, B. I., Colloquium: The Physics of Charge Inversion in Chemical and Biological Systems. *Rev. Mod. Phys.* **2002**, *74*, 329-345.

45. Jiang, J.; Cao, D.; Jiang, D.; Wu, J., Kinetic Charging Inversion in Ionic Liquid Electric Double Layers. *J. Phys. Chem. Lett.* **2014**, *5*, 2195-2200.

46. Israelachvili, J. N., *Intermolecular and Surface Forces*, 3rd ed.; Academic Press: Burlington, MA, 2011.

47. Israelachvili, J. N.; Pashley, R. M., Molecular Layering of Water at Surfaces and Origin of Repulsive Hydration Forces. *Nature* **1983**, *306*, 249-250.

48. Kalluri, R. K.; Konatham, D.; Striolo, A., Aqueous Nacl Solutions within Charged Carbon-Slit Pores: Partition Coefficients and Density Distributions from Molecular Dynamics Simulations. *J. Phys. Chem. C* **2011**, *115*, 13786-13795.

49. Messina, R.; Gonzalez-Tovar, E.; Lozada-Cassou, M.; Holm, C., Overcharging: The Crucial Role of Excluded Volume. *Europhys. Lett.* **2002**, *60*, 383-389.

50. Qiao, R.; Aluru, N. R., Ion Concentrations and Velocity Profiles in Nanochannel Electroosmotic Flows. *J. Chem. Phys.* **2003**, *118*, 4692-4701.

51. Feng, G.; Zhang, J. S.; Qiao, R., Microstructure and Capacitance of the Electrical Double Layers at the Interface of Ionic Liquids and Planar Electrodes. *J. Phys. Chem. C* **2009**, *113*, 4549-4559.

52. Feng, G.; Huang, J.; Sumpter, B. G.; Meunier, V.; Qiao, R., A "Counter-Charge Layer in Generalized Solvents" Framework for Electrical Double Layers in Neat and Hybrid Ionic Liquid Electrolytes. *Phys. Chem. Chem. Phys.* **2011**, *13*, 14723-14734.

53. Feng, G.; Qiao, R.; Huang, J.; Dai, S.; Sumpter, B. G.; Meunier, V., The Importance of Ion Size and Electrode Curvature on Electrical Double Layers in Ionic Liquids. *Phys. Chem. Chem. Phys.* **2010**, *13*, 1152-1161.

54. Jiang, D.; Wu, J., Unusual Effects of Solvent Polarity on Capacitance for Organic Electrolytes in a Nanoporous Electrode. *Nanoscale.* **2014**, *6*, 5545-5550.

TOC

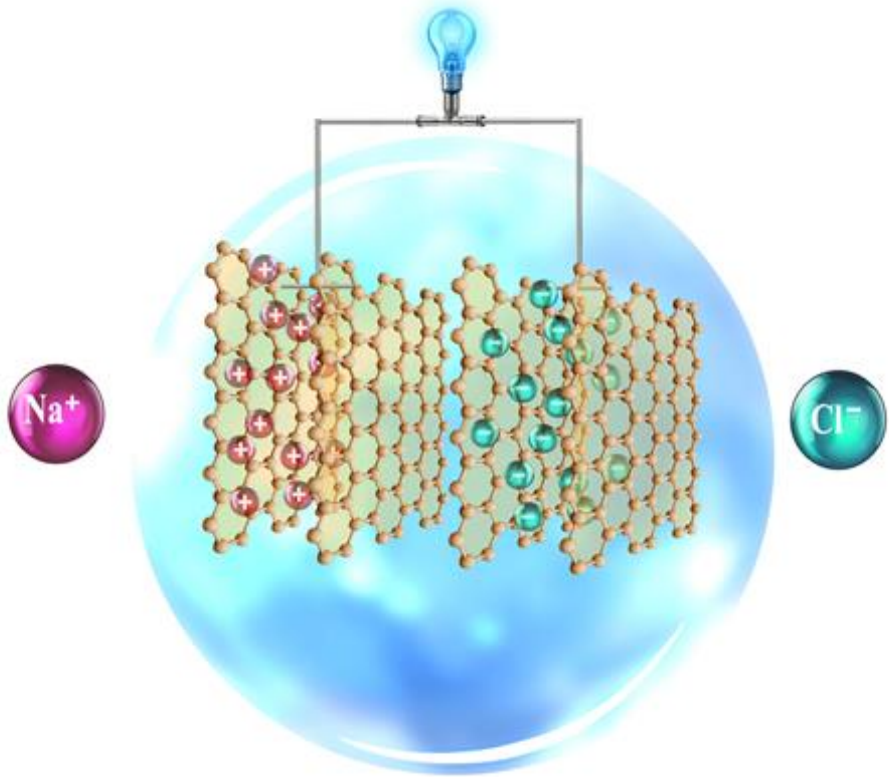